\documentclass[journal=jacsat,manuscript=article,layout=traditional,super=false]{achemso}
\setkeys{acs}{maxauthors=99, super=false}
\DeclareUnicodeCharacter{2212}{\textminus}
\usepackage{chemformula} 
\usepackage[T1]{fontenc} 
\setcitestyle{numbers,square} 

\makeatletter

\renewcommand{\@cite}[2]{[{#1\if@tempswa, #2\fi}]~}
\makeatother

\author{Patrizio Graziosi}
\affiliation{CNR − ISMN, Via P. Gobetti 101, Bologna 40129, Italy}
\author{Damiano Marian}
\affiliation{Dipartimento di Fisica, Università di Pisa, Largo Bruno Pontecorvo 3, Pisa 56127, Italy}
\email{damiano.marian@unipi.it}
\author{Andrea Tomadin}
\affiliation{Dipartimento di Fisica, Università di Pisa, Largo Bruno Pontecorvo 3, Pisa 56127, Italy}
\author{Stefano Roddaro}
\affiliation{Dipartimento di Fisica, Università di Pisa, Largo Bruno Pontecorvo 3, Pisa 56127, Italy}
\author{Omar Concepci\'on}
\affiliation{Peter Gruenberg Institute 9 (PGI-9) and JARA-Fundamentals of Future Information Technologies, Forschungszentrum Juelich, Juelich 52428, Germany}
\author{Johnny Tiscare$\mathrm{\tilde{n}}$o-Ram\'irez}
\affiliation{Peter Gruenberg Institute 9 (PGI-9) and JARA-Fundamentals of Future Information Technologies, Forschungszentrum Juelich, Juelich 52428, Germany}
\author{Agnieszka Anna Corley-Wiciak}
\affiliation{IHP - Leibniz Institute for High Performance Microelectronics, Frankfurt (Oder) 15236, Germany}
\altaffiliation{European Synchrotron Radiation Facility
71 avenue des Martyrs, CS 40220, Grenoble Cedex 9 38043, France}
\author{Dan Buca}
\affiliation{Peter Gruenberg Institute 9 (PGI-9) and JARA-Fundamentals of Future Information Technologies, Forschungszentrum Juelich, Juelich 52428, Germany}
\author{Giovanni Capellini}
\affiliation{IHP - Leibniz Institute for High Performance Microelectronics, Frankfurt (Oder) 15236, Germany}
\altaffiliation{Dipartimento di Scienze, Università degli Studi Roma Tre, Viale G. Marconi 446, Roma 00146, Italy}
\author{Michele Virgilio}
\affiliation{Dipartimento di Fisica, Università di Pisa, Largo Bruno Pontecorvo 3, Pisa 56127, Italy}

\title{Epitaxial SiGeSn alloys for CMOS-compatible thermoelectric devices}


\begin{document}

\begin{abstract}
The integration of thermoelectric devices into mainstream microelectronic technological platform could be a major breakthrough in various fields within the \emph{so-called} Green-IT realm. In this article, the thermoelectric properties of heteroepitaxial SiGeSn alloys, a novel CMOS compatible material system, are evaluated to assess their possible application in thermoelectric devices. To this purpose, 
starting from the experimentally low lattice thermal conductivity of SiGeSn/Ge/Si layers of about $\sim$1-2  W/m$\cdot$K assessed by means of 3-$\omega$ measurements, the figure of merits are calculated through the use of Boltzmann transport equation, taking into account the relevant inter-valley scattering processes, peculiar of this multi-valley material system. Values for the figure of merit $ZT$ exceeding $1$ have been obtained for both p- and n- type material at operating temperatures within the 300---400 K range, i.e. at a typical On-Chip temperatures. In this interval, the predicted power factor also features very competitive values of the order of 20 $\rm{\mu W/cm\cdot K^2}$. Our finding indicates that this new class of Si-based materials has extremely good prospects for real-world applications, and can further stimulate scientific investigation in this ambit. 
\end{abstract}

\maketitle

\section{Introduction}\label{sec1}
Thermoelectric (TE) devices provide a solution for direct energy conversion for applications and technologies that require a reliable source, rather than an efficient one~\cite{Champier2017}.  Thermoelectric generators (TEG) are used to convert low-grade heat in energy to power systems and devices for a variety of applications~\cite{Fernandez-Yanez2021} as e.g. in wearable health monitoring~\cite{Zadan2024}, automotive~\cite{Quan2024}  and  aerospace technologies~\cite{Cataldo2011},  and manufacturing~\cite{Kuroki2015}.  Despite these achievements,  the diffusion of TEG remains limited in large consumer markets due to major drawbacks of TE technologies available to date~\cite{Yan2022}: i) low conversion efficiency -especially for room temperature operations-;  ii) challenges in TEG miniaturization and integration in mass production manufacturing, leading to relatively high cost per device; iii) TEG generator operating at room temperature (RT) are usually based on  toxic materials or materials of limited availability~\cite{Cao2023}. 

The potential performance of a TE material is  usually quantified by the  dimensionless figure of merit~\cite{Kim2015} $ZT=\frac{S^2\sigma T}{\kappa}$, where $S$ is the Seebeck coefficient and $\sigma$ and $\kappa$ are the electrical and thermal conductivity, respectively. A 300 K value of $ZT$ $\approx$ 1 is required for realizing a device with an efficiency of practical use (about 15$\%$ of the Carnot limit).
An ideal TE material should then feature a low $\kappa$, needed to preserve the T gradient across the device, while maintaining excellent electrical transport properties. However, there are well known constraints to the optimization of the material parameters. As a matter of fact, $\sigma$ in semiconductors can be increased by leveraging on the doping concentration, but as dictated by the Wiedemann-Franz law, this also induces larger $\kappa$ values, due to the simultaneous increase of the electronic contribution $\kappa_e$ to the thermal conductivity, thus imposing a trade-off to the optimal carrier density~\cite{Wang2019}.

\begin{figure}[h!!!!!!]
\centering
\includegraphics[width=0.9\columnwidth]{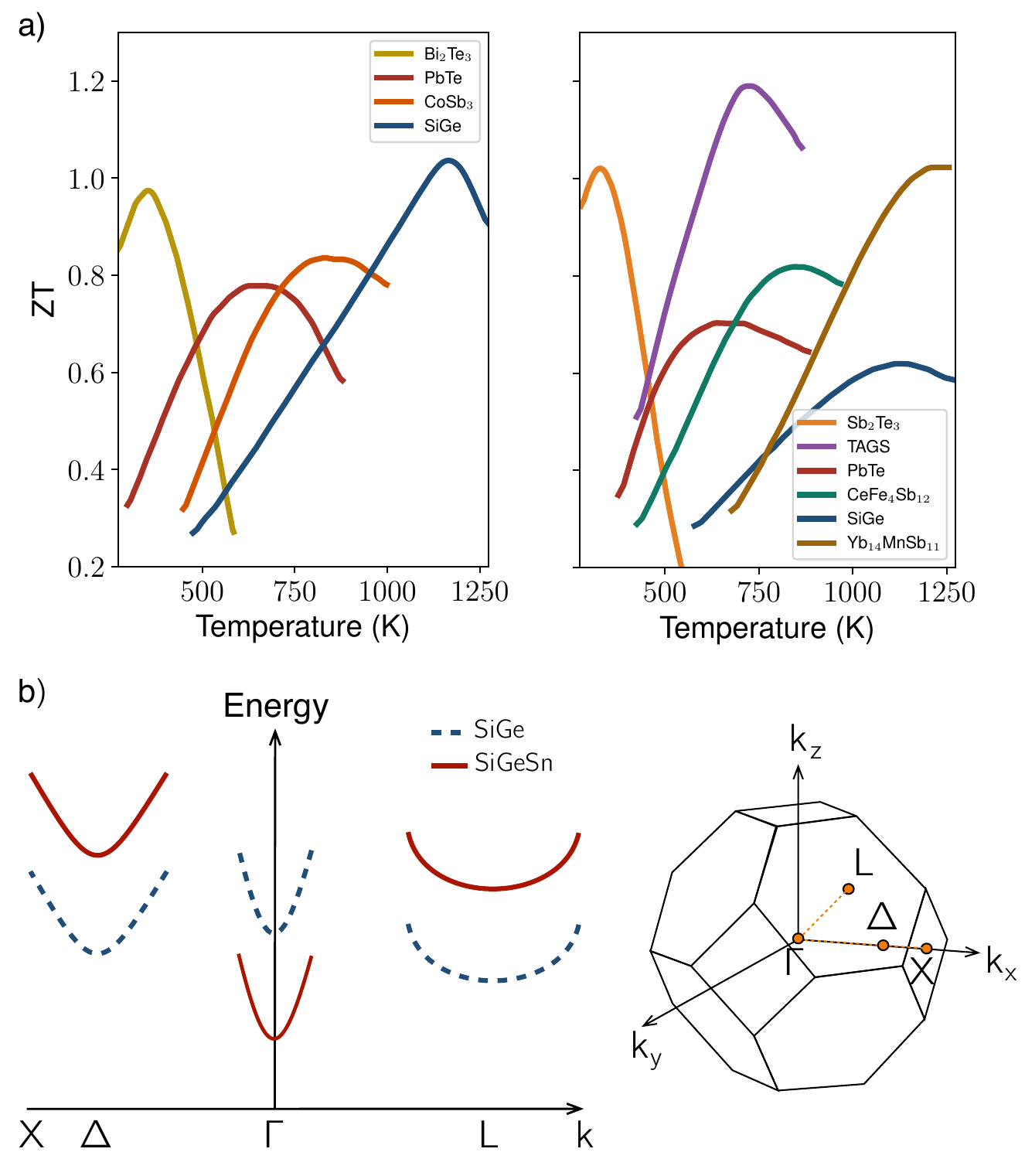}
\caption{a) $ZT$ figure of merit as a function of temperature for different semiconductor n-type (left) and p-type (right) materials. b) Left: schematic conduction band structure of a Ge-rich SiGe alloy (blue dashed). Alloying with Sn induces the \emph{relative} downshift of the $\Gamma_c$ band edge (solid curve) with respect to the $\Delta$ and L ones. L, $\Gamma_c$ and $\Delta$ point conduction minima are typically found within a 100-200 meV energy range. Right: first Brillouin zone showing the position of these high symmetry points. }
\label{fig1}
\end{figure}

Si, Ge, and their alloys are CMOS-compatible materials and have been used in high temperature TEG devices having a high $ZT$ values peak at around 1200 K, see Fig.~\ref{fig1}(a). However, their $ZT$ dramatically drops at 300 K, also in nanostructurated devices, engineered in the past decades in an attempt to enhance the TE performance of the bulk material ~\cite{Dresselhaus2007,Li2003,Yu2012}. For this reason, today SiGe alloys are employed only at elevated temperatures in niche applications, as e.g. for radioisotope thermoelectric generators in space missions. On the other hand, commonly used TE materials, such as Bi$_2$Te$_3$ and PbTe, exhibit high $ZT$ values at room temperature (Fig.~\ref{fig1}(a)) due to the their low thermal conductivity, but their toxicity and incompatibility with silicon-based microelectronics standards prevent their widespread use in real-world applications~\cite{Shi2025}. 
Consequently, in view of potential large-scale applications integrated with consumer electronics, TE materials should be based on an innovative, CMOS-compatible, and non-toxic material system, having thermal conductivity at 300 K suppressed by efficient phonon-phonon scattering processes, but still characterized by high charge carrier conductivity.

Ge-rich SiGeSn ternary alloys have recently emerged as a transformative Group IV material system. As a matter of fact, the addition of Sn into the SiGe matrix introduces significant changes in the material's electronic and thermal properties since Sn incorporation: $i)$ dramatically reduces the lattice thermal conductivity due to the high mass contrast with the Si and Ge ions; $ii)$ can induce a down-shift of the $\Gamma$ conduction valley below the energy of the L-point minimum, thus realizing a direct bandgap semiconductor material as schematically depicted in Fig.~\ref{fig1}(b). 
Consequently, in n-type systems, larger electron mobilities with respect to SiGe can be obtained. 
The rationale for this mobility enhancement is related to the proximity of the $\Delta$-, L- and $\Gamma$-point band edges, typically laying within a $\sim$100 meV range in this multi-valley quasi-direct semiconductor. It follows that, alloying with Sn induces a significative increase of the carrier population which populates the $\Gamma$-valley, which features a lighter transport mass with respect to the L and $\Delta$ ones~\cite{MA_alloy}. Furthermore, the non-polar character of SiGeSn alloys contributes to reduce the thermally induced suppression of the electrical conductivity, due to the absence of Fr\"ohlich interaction.

The Si$_{x}$Ge$_{1-x-y}$Sn$_{y}$ material system has come under the spotlight thanks to the recent advancements in the epitaxial growth of high-quality Ge-rich SiGeSn layers on Si-based Ge/Si virtual substrates, using tools and processes compatible with the standards of the Si-CMOS microelectronics industry~\cite{Kuroki2015}. In particular, the possibility of an extensive band gap engineering through composition has attracted the silicon photonics community~\cite{Moutanabbir2021}.  As a matter of fact, the so-called ``directness'' in Ge-rich alloys, obtained by Sn incorporation as stated above, has a huge positive impact on the radiative recombination efficiency of the material, leading to the demonstration of optically pumped GeSn-based laser operating at room temperature~\cite{Buca2022} or cw- electrically pumped laser based on SiGeSn/GeSn multi quantum wells~\cite{Seidel2024}.
The directness is also exploited in advanced microelectronic devices such as Tunnel-FET where the direct band-gap increases the device performances thanks to more efficient  tunneling rates of the $\Gamma$-valley electrons which, for momentum conservation reasons, do not require phonon assistance~\cite{Liu2022}. 
Finally, since the optimization of $ZT$ usually requires high carrier densities, it is important to notice that, thanks to the aforementioned recent advances, doping  of Ge-rich SiGeSn layers with active impurity concentrations exceeding $10^{20}$~cm$^{-3}$ has been achieved for both p-type and n-type systems~\cite{Frauenrath2023}. 

It is clear that the unequivocal demonstration of high SiGeSn TE performances around 300 K could pave the way to a multifunctional and integrated platform, comprising electronic, photonic, and TE devices, manufactured within the Si-CMOS standard, with all the inherent advantages granted by the microelectronic technology. 
This multifunctional toolbox would greatly contribute to the \emph{so-called} green-IT objectives. In fact, due to its scalability, environmental friendliness, and low-cost, it can be employed in large-scale devices designed for energy-harvesting on-chip~\cite{Zhang2023} or chip cooling~\cite{CHEN2022100700}, temperature-tunable photonic circuits for optical computation, sensing, and lab-on-chip architectures. 

In this context, given the very promising low lattice thermal conductivity $\kappa_{l}$ observed in the last few years for GeSn~\cite{Spirito2021,ConcepcionGeSn2023,ConcepcionGeSn2024}, studies have emerged in the literature targeting the numerical prediction of $\kappa_{l}$ also in Si$_{x}$Ge$_{1-x-y}$Sn$_{y}$ alloys, in the entire~\cite{Khatami2016} or for a selected subset~\cite{Lee2017} of the $(x,y)$ parameter space. 
These findings are very promising for TE applications since suggest a very robust suppression of $\kappa_{l}$ with respect to SiGe and GeSn, with values as low as 1 W/m$\cdot$K. Here, we move a step further in this direction combining lattice thermal conductivity measurements with a comprehensive numerical investigation of the electronic transport properties of the Ge-rich Si$_{x}$Ge$_{1-x-y}$Sn$_{y}$ heteroepitaxial alloys. In this way, we are able to assess the TE properties of the ternary material system at 300 and 400 K, in the entire portion of the $(x,y)$ parameter region experimentally accessible. Indeed, owing to the very low solid solubility of Sn in both Si and Ge, only tin content below $\approx$0.2 can be obtained without compromising on the electrical transport properties~\cite{Grutzmacher2023}.

We first performed 3-$\omega$ experiments to measure the lattice conductivity of a Ge-rich Si$_{x}$Ge$_{1-x-y}$Sn$_{y}$ sample set featuring different Sn concentrations. Subsequently, upon combining these data with literature values of $\kappa_{l}$,  we achieved an estimation of lattice thermal conductivity across the entire $(x,y)$ parameter region.
As for the electronic transport properties and their contribution to the thermal conductivity, we relied on the Boltzmann transport equation, beyond the constant time relaxation approximation, as implemented in the ElecTra simulation suite~\cite{ELECTRA,ELECTRA_GitHub}, where particular care has been devoted to include all the intervalley scattering processes, peculiar of this multivalley material system, as well as bipolar effects which are active at around 300 K due to the relatively small value of the band gap. 
After validating the ElecTra results against experimental mobility data for GeSn, we systematically explore the alloy parameter space of the ternary material to optimize its $ZT$ values at different temperatures. 

Our findings suggest that p- and n-type SiGeSn alloys can achieve $ZT$ values greater than 1 in the 300 K to 400 K temperature range, a performance comparable to commercial TE materials. This significant result demonstrates the potential of SiGeSn to bridge the gap between high-temperature TE materials and room-temperature applications, offering a sustainable, non-toxic, and CMOS-compatible solution for next-generation thermoelectric devices.

\section{Methods and validation}
\subsection{Electronic transport properties}
To estimate the figure of merit $ZT$ and the power factor PF for p- and n-type Si$_{x}$Ge$_{1-x-y}$Sn$_{y}$ alloys, we assessed their charge transport properties and the electronic contribution to the thermal conductivity $\kappa_{e}$ by means of numerical simulations based on the Boltzmann transport equation, which allowed us to explore the parameter space spanned by the Si and Sn concentration, the doping density, and the lattice temperature. 
Due to the quasi-direct character of the SiGeSn material system, we have taken into account the carrier population in the different conduction valleys, which correspond to the minima at the $\Gamma_c$, L, and $\Delta$ point of the Brillouin zone, as well as their interaction through inter-valley scattering events.
For this purpose, we relayed on the ElecTra code~\cite{ELECTRA,ELECTRA_GitHub}, a well-established simulation suite, developed by one of the Authors, which in the last few years has been extensively tested with group IV elemental materials and SiGe alloys. 

ElecTra offers the functionality of charge transport calculation beyond the constant relaxation time approximation by considering $i$) the full energy/momentum/band dependence of the scattering rates, $ii$) the distinction between intra- and inter-band transitions, and $iii$) the bipolar transport resulting from the joint contribution stemming from the valence and conduction carriers. Moreover, to increase the accuracy, calculations are performed in a full-band approach and using anisotropic scattering rates.
The scattering mechanisms taken into account are the electron-phonon coupling with the acoustic and optical branches, the Coulomb interaction induced by charged donors/acceptors ions, and the alloy disorder potential.

Due to the ternary and multivalley character of the investigated material system, particular care has been devoted to upgrade the description of the alloy scattering, since literature models deal mainly with single valley and/or III-V semiconductors.
To this aim, as first step, the square modulus of the alloy scattering matrix element for Si$_{x}$Ge$_{1-x-y}$Sn$_{y}$ has been linearly decomposed, as outlined in Ref.~\cite{FerryAlloy78}, in terms of the binary SiGe and GeSn ones, weighted according to the Si:Sn ratio in the ternary material. 
Moreover, as outlined in Ref.~\cite{Klimeck_alloy} for the SiGe case, to evaluate each binary scattering matrix element, we have distinguished between intra- and inter-valley events, adopting specific coupling potentials whose values depend on the initial and final valley state. 
The alloy scattering potentials used in our calculations for SiGe are taken from the theoretical estimations obtained  by means of tight-binding supercell calculations in Ref.~\cite{Klimeck_alloy}, where also a validation procedure is discussed. Indeed, the Authors of Ref.~\cite{Klimeck_alloy} demonstrate that using their alloy scattering parameters, the experimental mobility of SiGe is successfully reproduced in the entire compositional range.
For the case of GeSn, we relied on the theoretical values resulting from the supercell simulations based on the density functional theory reported in Ref.~\cite{MA_alloy}. However, after the model calibration procedure discussed hereafter, a rescaling factor has been applied to the set of GeSn alloy potentials for the conduction band.
As for the electron-phonon coupling, the adopted phonon energies and scattering parameters are reported and discussed in the SI, where we also list the values for all the other material parameters used in our simulations.

The aforementioned model calibration procedure has been performed using a set of high-quality p-and n-type GeSn layers, featuring a thickness between 250 and 300 nm and different Sn concentrations spanning the 5-15\% interval. 
These samples were grown on 200 mm Si(100) wafers using an industry-compatible reduced-pressure chemical vapor deposition reactor. In order to enhance the crystal quality, a 300 nm thick Ge buffer layer was introduced to accommodate the significant lattice mismatch between Sn and Si. 

\begin{figure}[h!!!]
\centering
\includegraphics[width=0.99\columnwidth]{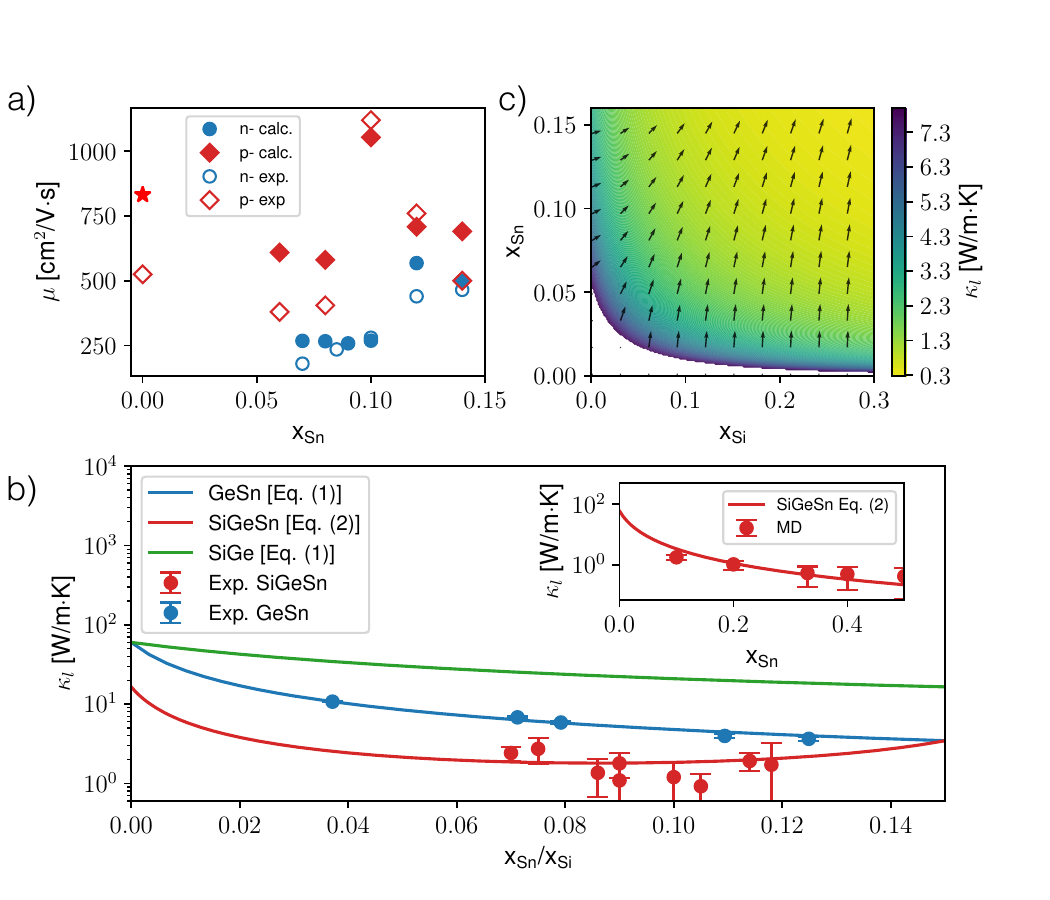}
\caption{a) GeSn electron (blue) and hole (red) mobilities as a function of Sn content, measured with Hall bar (empty symbols) and theoretically predicted by ElecTra (full symbols). For comparison we also reported the mobility of bulk Ge at the same p-type impurity concentration of our Ge sample (red star). b) Lattice thermal conductivity from 3-$\omega$ experiments for Ge$_{1-x}$Sn$_x$ (blue points) and Si$_{x}$Ge$_{1-x-y}$Sn$_y$ with $x+y=0.15$ (red points). Data are shown as a function of the Sn concentration together with the estimation based on Eq.~(\ref{k-binary}) and Eq.~(\ref{k-ternary}). For comparison, $\kappa_l$ of SiGe (green curve) as a function of the Si concentration is also reported. In the inset molecular dynamics (MD) theoretical conductivity data (red points) from Ref.~\cite{Lee2017} are compared with the estimation based on Eq.~(\ref{k-ternary}). c) Colour map of the lattice thermal conductivity of Si$_{x}$Ge$_{1-x-y}$Sn$_y$ for Si and Sn contents in [0-0.3] and [0-0.16] respectively; black arrows represent the gradient field.}
\label{fig2}
\end{figure}

N-type epilayers have been in-situ doped by co-deposition of Phosphorous, achieving carrier densities ranging in the 2.4---3.8 $\times10^{19}$ cm$^{-3}$, as probed by means of Hall measurements at 300 K. From Hall experiments, performed on nominally intrinsic samples, we also asses p-type carrier densities  the varying in the $10^{16}$---$10^{18}$ cm$^{-3}$ range, depending on the Sn content. Details on the epitaxy and structural characterization of the samples can be found in Refs.~\cite{ConcepcionGeSn2023,ConcepcionGeSn2024}. 
Carrier mobilities are shown in Fig.~\ref{fig2}(a) as a function of the Sn concentration. In the p-type material, mobility vs Sn content does not exhibit a clear trend. As a matter of fact, in this case the scattered values of $\mu$ simply reflect the differences in the doping concentrations, with a peak value around 1100 $\mathrm{cm^2/V\cdot s}$, obtained for the Ge$_{90}$Sn$_{10}$ sample since thanks to its lowest carrier density (1.2$\times$10$^{16}$cm$^{-3}$).
Conversely, in the n-type sample set, we observe for Sn concentration in the 7-14 \% interval a well-defined monotonic increase of $\mu$, from $\sim$ 230 to 465 $\mathrm{cm^2/V\cdot s}$, due to the quite homogeneous doping contents.
This result is in semi-quantitative agreement with the theoretical predictions reported in Ref.~\cite{MA_alloy}, and has to be attributed to the enhancement of the carrier population in the $\Gamma_c$ valley, triggered by the lowering of the $\Gamma_c$-L energy barrier. As a matter of fact, despite the increased alloy scattering rates, the much lower effective mass of the $\Gamma_c$ electrons, with respect to the L ones, positively impacts mobility, which is expected to exceed the one of Ge when the Sn content is above 15\%~\cite{MA_alloy}.

Using the potential values reported in Ref.~\cite{MA_alloy} we obtained a nice agreement with the measured mobility for the p-type materials only, while our theoretical predictions for the electron mobility overestimated the experimental data. 
Therefore, to calibrate the model, the alloy scattering potentials predicted for the conduction band have been tuned to match the experimental data, achieving a quantitative consistency (see Fig.~\ref{fig2}(a)) when a 0.7 scaling factor is applied.

After this calibration procedure, we separately inspected the different channel contributions which limit the carrier mobility. 
In the n-type material, up to 10\% of Sn content, the alloy scattering is the dominant one, representing $\sim$ 60\% of the total scattering rate and about the 90\% of its lattice part. At larger Sn concentrations, the relative weight of the alloy scattering decreases in favor of the electron-phonon interaction due to band-structure effects related to the energy separation between the $\Gamma_c$ and L valley.  
For the p-type case, we find that at any composition the alloy scattering rate is  comparable to the electron-phonon one. However, in the low Sn concentration regime their impact on the total mobility is modest, while it increases up to nearly 50\% at the higher Sn concentrations. 

\subsection{Lattice thermal conductivity of SiGeSn}
Now, we focus on the assessment of the lattice thermal conductivity in the SiGeSn material system. To this aim, we have performed 3-$\omega$ experiments~\cite{Cahill90} on a set of Si$_x$Ge$_{1-x-y}$Sn$_y$ layers with $x\in[0.07-0.12]$.
The investigated samples, grown on a Ge buffer in the same reactor presented above, feature a constant Ge content $1-x-y=0.85$ and thicknesses ranging in the 40 --- 360 nm interval. 
Our estimates for $\kappa_l$ as a function of the Sn content $y$ are shown as red points in Fig.~\ref{fig2}(b) together with the GeSn lattice thermal conductivity measured in Ref.~\cite{ConcepcionGeSn2024} for Sn contents in a similar range (blue points).
As further  discussed in the following, it is apparent from Fig.~\ref{fig2}(b) that alloying Ge with both Si and Sn induces a rapid and stronger suppression of $k_l$ with respect to the GeSn binary system.

Combining our 3-$\omega$ data with literature experimental~\cite{ConcepcionGeSn2024} and theoretical~\cite{Lee2017} estimates, we were able through the fitting procedure detailed below to propose a phenomenological relation which describes the lattice thermal conductivity of the Si$_x$Ge$_{1-x-y}$Sn$_y$ ternary system in the sub-set of the compositional parameter $(x,y)$ space defined by $x\in[0-0.3]$ and $y\in[0-0.16]$.
The rationale underlying the choice of this parameter region is two-fold. On one side Sn rich alloys become thermodynamically unstable for Sn concentration greater than $\sim0.16$. On the other side, TE properties of Ge-rich materials are expected to over-perform with respect to their Si-rich counterpart, due to the lower lattice thermal conductivity of Germanium. Furthermore, we prefer not to provide TE estimates at $(x,y)$ regions located too far from the currently available $\kappa_{l}$ data. 

As a starting point, we modeled $\kappa_l$ of the Si$_{x}$Ge$_{1-x}$ and Ge$_{1-x}$Sn$_{x}$ binary alloys in terms of the thermal conductivity of their elemental constituents and of a fitting parameter $A$, relying on the following equation~\cite{Wagner2006}:

\begin{equation}
    \kappa_{\text{SiGe/GeSn}}(x) = \left(\frac{x}{\kappa_{\text{Si/Sn}}} + \frac{1-x}{\kappa_{\text{Ge}}} + \frac{(1-x)x}{A_{\text{SiGe}}/A_{\text{GeSn}}}\right)^{-1}.
    \label{k-binary}
\end{equation}

The functional form of the above relation for $\kappa_l(x)$ corresponds to a U-shaped curve when $x\in[0,1]$. For SiGe alloys, following Ref.~\cite{Wagner2006} we set in Eq.~(\ref{k-binary}) $\kappa_{\text{Si}} = 148$ W/m$\cdot$K, $\kappa_{\text{Ge}} = 60 $ W/m$\cdot$K and $A_{\text{SiGe}} = 2.8$ W/m$\cdot$K, which corresponds to the green curve shown in Fig.~\ref{fig2}(b). For the GeSn case, after fixing $\kappa_{\text{Sn}} = 63$ W/m$\cdot$K, we have fitted the data reported in Ref.~\cite{ConcepcionGeSn2024}, obtaining  $A_{\text{GeSn}} = 0.467$ W/m$\cdot$K (blue  curve in Fig.~\ref{fig2}(b)). 
It is apparent from Fig.~\ref{fig2}(b) that when alloying Ge, Sn is much more effective than Si in suppressing the lattice conductivity. Indeed, Eq.~(\ref{k-binary}) indicates that already at 10\% of Sn content, $\kappa_{\text{GeSn}}$ drops at 5 W/m$\cdot$K, i.e. a factor of 4 lower than the $\kappa_{\text{SiGe}}$ conductivity at the same Si content.

Next, we describe the thermal conductivity of the ternary Si$_{x}$Ge$_{1-x-y}$Sn$_{y}$ alloy relying on a phenomenological generalization of the above relation, defined for $x + y \leq 1$:
 
\begin{equation}
    \kappa_{\text{SiGeSn}}(x, y) = \left(\frac{x}{\kappa_{\text{Si}}} + \frac{y}{\kappa_{\text{Sn}}} + \frac{1-x-y}{\kappa_{\text{Ge}}} + \frac{x(1-x)}{A_{\text{SiGe}}} + \frac{y(1-y)}{A_{\text{GeSn}}} + \frac{xy}{A_{\text{SiGeSn}}}\right)^{-1}.
    \label{k-ternary}
\end{equation}
Eq.~(\ref{k-ternary}) reproduces Eq.~(\ref{k-binary}), for the binary alloys SiGe and GeSn, i.e. when $y = 0$ and $x = 0$, respectively.
In the above equation, we have introduced a term proportional to the product of the Si ($x$) and Sn ($y$) concentration, controlled by $A_{\text{SiGeSn}}$ which plays the role of a new fitting parameter.
Our best estimate for $A_{\text{SiGeSn}}$ gives $0.016$ W/m$\cdot$K. This value has been obtained simultaneously fitting with Eq.~(\ref{k-ternary}) the 3-$\omega$ measurements shown in Fig.~\ref{fig2}(b) for $x+y=0.15$ and the $\kappa_l$ theoretical estimations provided in Ref.~\cite{Lee2017} for $x=y$ with $x\in[0.1,0.5]$, which has been obtained by molecular dynamics (MD) simulations. 

The resulting $\kappa_l$, plotted as a function of the Sn content is shown as red curve in Fig.~\ref{fig2}(b) for the $x+y=0.15$  case, while comparison of Eq.~(\ref{k-ternary}) with theoretical MD points from Ref.~\cite{Lee2017} at $x=y$ is reported in the inset.
Beside noticing the effectiveness of the adopted fitting formula in describing $\kappa_l$ of the ternary system in the Ge-rich domain, we stress again that the addition of Si induces, already for modest concentrations, a stronger suppression of the Ge thermal conductivity with respect to the GeSn material. As a matter of fact, typical $\kappa_l$ values in Fig.~\ref{fig2}(b) are around 2 W/m$\cdot$K i.e. a factor of 2.5 lower with respect to the GeSn ones reported in Ref.~\cite{ConcepcionGeSn2024} for Sn content $\approx$10\%. 
$\kappa_l$ as a function of the Si and Sn concentrations with $x\leq0.3$ and $y\leq 0.16$ estimated through Eq.~(\ref{k-ternary}) is shown in Fig.~\ref{fig2}(c). In the plotted domain, we observe a bowl-shape behavior with strong gradients close to the domain boundaries, pointing approximately toward the main diagonal of the $(x,y)$ plane, where $\kappa_l$ reaches the lowest values, with an absolute minimum as low as $0.28$ W/m$\cdot$K, obtained for Si$_{0.3}$Ge$_{0.54}$Sn$_{0.16}$.

\begin{figure}[h!!!]
\centering
\includegraphics[width=0.8\columnwidth]{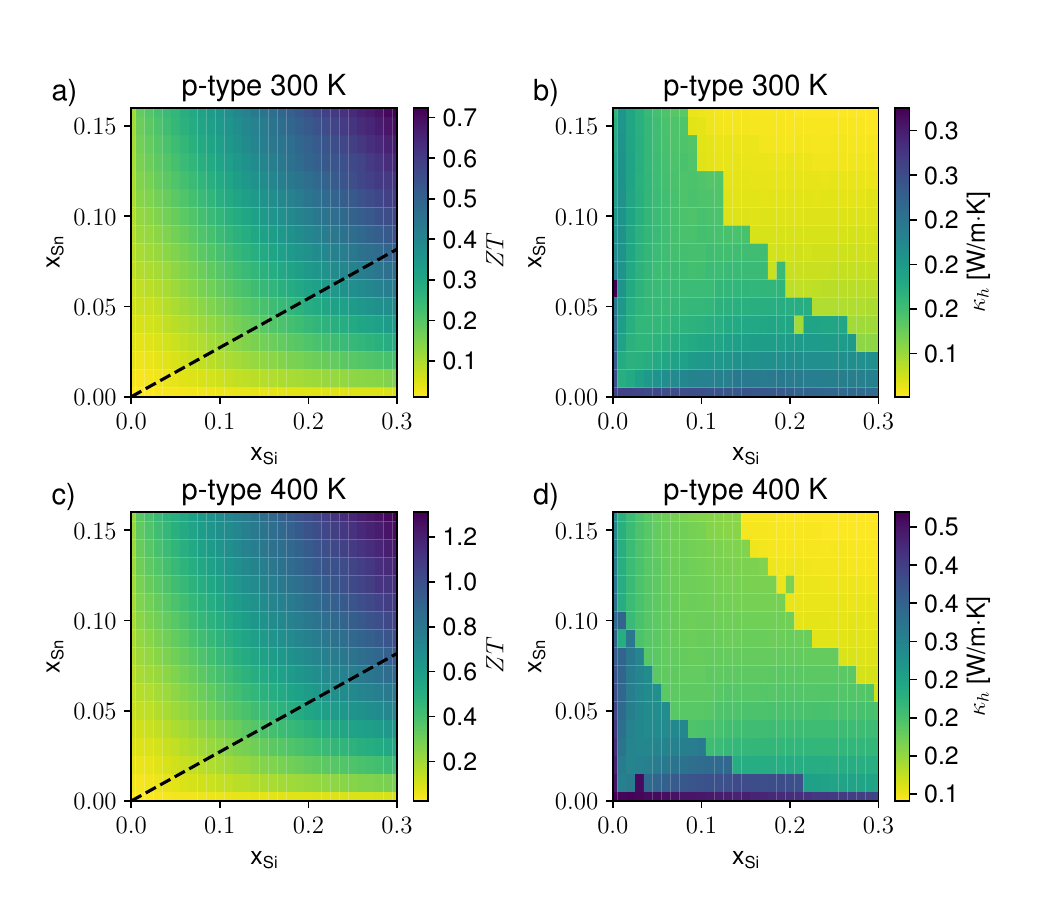}
\caption{Color map of the $ZT$ figure of merit, predicted for p-type SiGeSn at the optimal carrier concentration at 300 K (a) and 400 K (c). The corresponding hole contribution  to the thermal conductivity $\kappa_h$ is reported in panel (b) and (d). The black dashed curve in panels (a) and (c) evidences the lattice matching condition with the relaxed Ge buffer material.}
\label{fig3}
\end{figure}

\section{TE performance and discussion}
Combining the fit values of the lattice thermal conductivity with the electronic properties predicted by ElecTra, we obtained the $ZT$ and PF figures of merit at 300 K and 400 K. 
Our $ZT$ predictions for the p-type material at the optimal hole density are reported in Fig.~\ref{fig3}(a) where the black dashed curve evidences the x:y ratio equal to 3.67. 
This value corresponds to a Si$_{x}$Ge$_{1-x-y}$Sn$_{y}$ lattice constant matching the one of Ge, thus indicating the absence of biaxial deformation when a relaxed Ge buffer layer is introduced to accommodate the mismatch between the Si and SiGeSn lattice parameter.
 $ZT$ increases at larger Sn and Si contents, since its qualitative behavior is controlled by $\kappa_l(x,y)$.
As a matter of fact, differently from what holds for the conduction band, alloying with Si and Ge does not induce complex band-structure effects in the valence, which maintains its single valley character. It follows that the trend observed in Fig.~\ref{fig3}(a) is triggered by the suppression of $\kappa_l(x,y)$, which governs also the functional form of the hole contribution $\kappa_{h}(x,y)$ (Fig.~\ref{fig3}(b)) to the total thermal conductivity and of the optimal hole density $p_{opt}(x,y)$, plotted in the SI.
In the explored domain $p_{opt}$ varies in the 4 --- 10$\times$10$^{18}$~cm$^{-3}$ interval, which corresponds to $\kappa_{h}$ values in the 0.1 --- 0.35 W/m$\cdot$K range. 
The lowest $p_{opt}$ concentrations are expected in the Sn- and Si-rich region, where the strong suppression of $\kappa_l$ poses an upper bound to $\kappa_{h}$ when maximizing $ZT$.

Spanning the $(x,y)$ parameter space, we find that $ZT$ peaks for Si$_{0.3}$Ge$_{0.54}$Sn$_{0.16}$ at 0.7, a value 6 times larger than the one calculated in Ref.~\cite{ConcepcionGeSn2024} for the binary GeSn alloy at 0.15\% Sn content.
Remarkably, a modest increase of the lattice temperature triggers even larger $ZT$ values as apparent from the bottom panels of Fig.~\ref{fig3}.
Indeed, having in mind energy harvesting applications in the field of microelectronics leveraging on this CMOS compatible material, we have calculated TE properties also at 400 K, a typical chip operation temperature, predicting in this case $ZT$ values as high as 1.3 at a hole density of about 5$\times$10$^{18}$~cm$^{-3}$ for Sn and Si contents of 0.16 and 0.3, respectively.

\begin{figure}[h!!!]
\centering
\includegraphics[width=0.8\columnwidth]{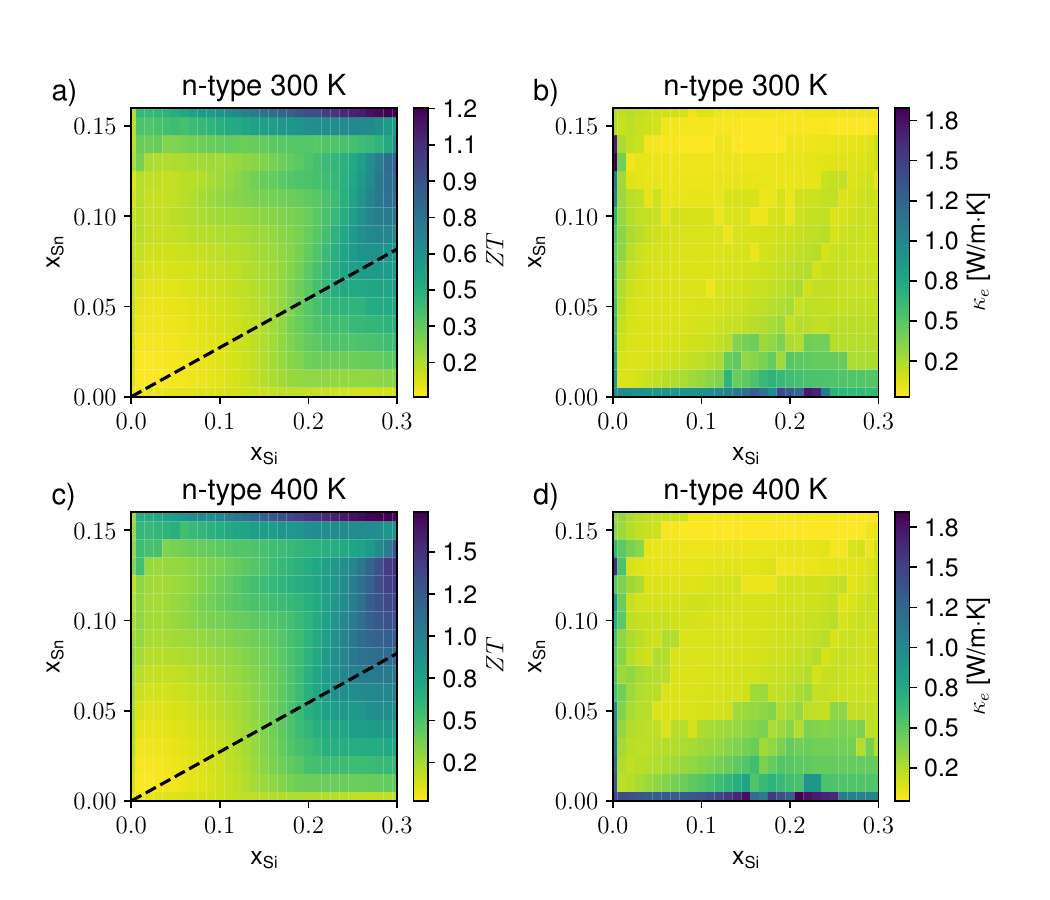}
\caption{Color map of the $ZT$ figure of merit, predicted for n-type SiGeSn at the optimal carrier concentration at 300 K (a) and 400 K (c). The corresponding electron contribution  to the thermal conductivity $\kappa_e$ is reported in panels (b) and (d). The black dashed curve in panels (a) and (c) evidences the lattice matching condition with the relaxed Ge buffer material.}
\label{fig4}
\end{figure}

We now focus on the n-type material whose $ZT$ figure of merit is shown in Fig.~\ref{fig4} as a function of $(x,y)$ in the same subset of the alloy parameter space. Top and bottom panels refers to 300 and 400 K, respectively and as for the p-case, the right panels display the electronic contribution $\kappa_e$ to the total thermal conductivity.
The qualitative behavior of $ZT$ differs from the one discussed above, indicating that alloy induced band-structure effects in the conduction, play a relevant role, due to their non-trivial impact on the electronic conductivity.
In particular we expect a complex functional dependence of the effective mobility,  arising from the joint action of the content dependent alloy scattering rate and of the electron density of $\Gamma_c$ carriers.
As a matter of fact, as mentioned in the introduction, the conductivity mass of $\Gamma_c$ electrons is much lighter than the one associated with the $L$ valley and the ratio between the two populations is greatly influenced by the Sn and Si content.
As a result, we find that the largest $ZT$ are still found at the highest Si concentration, however in this case their values do not increase monotonically with the Sn content. 
As a matter of fact, we observe in the color maps of Fig.~\ref{fig4}a) and c) two regions of interest, one at high Sn content, and another one at high Si and lower Sn contents. The latter is of special interest since Si has infinite miscibility in Ge whereas Sn has limited solid-state miscibility in Si and Ge~\cite{Grutzmacher2023}. 
In particular, at the $x$ Si concentration of 0.3 we predict a peak $ZT$ equal to 1.2 (1.6) at 300 K (400 K), achieved for $y$ Sn content of 0.16 while for $y$ = 0.13 we get $ZT$ = 0.85 (1.5).  
We also notice that for the n-type material the optimal doping densities are larger, covering the 5 --- 7 $\times$10$^{19}$ cm$^{-3}$ range (see SI), which however remains experimentally accessible~\cite{Frauenrath_2022}.

\begin{figure}[h!!!]
\centering
\includegraphics[width=0.8\columnwidth]{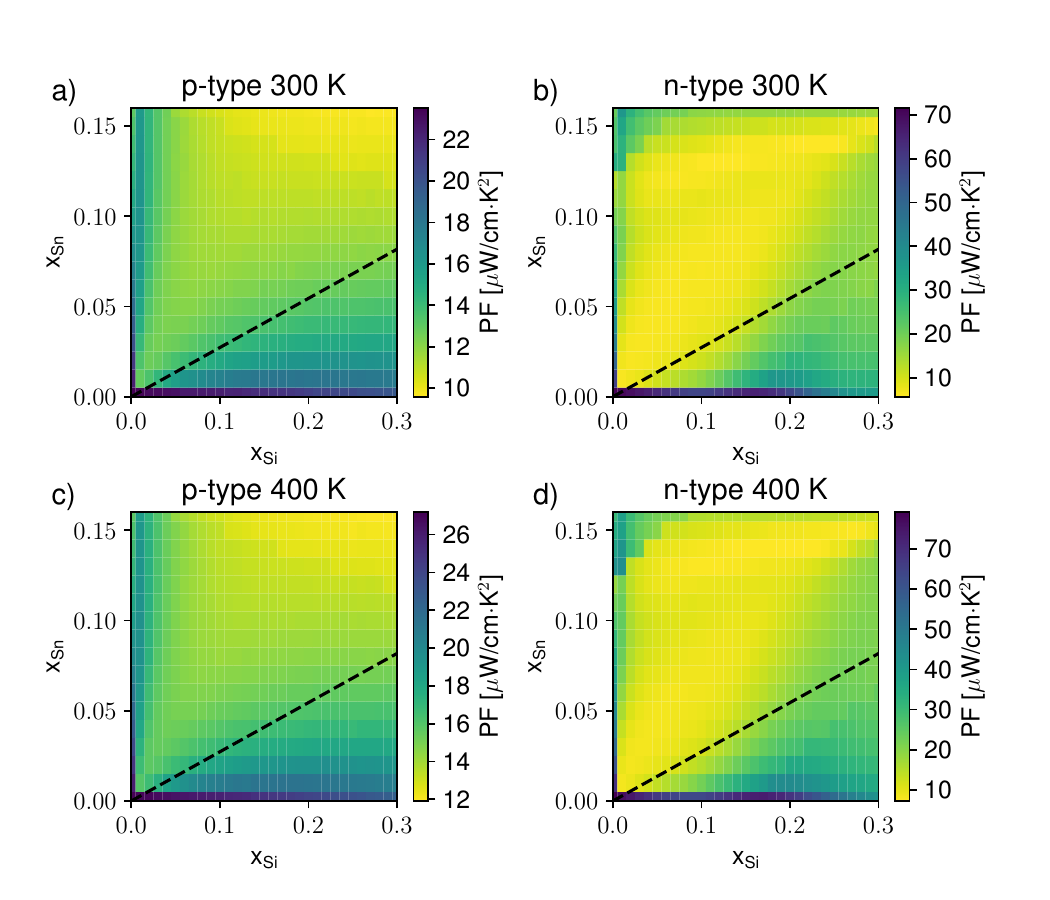}
\caption{Color map of the Power Factor PF at 300 K at the optimal carrier density for p-type (a) and n-type (b) SiGeSn alloys. Corresponding quantities calculated at 400 K are shown in panel (c) and (d). The black dashed curve evidences the lattice matching condition with the relaxed Ge buffer material.}
\label{fig5}
\end{figure}

In Fig.~\ref{fig5} we show 300 and 400 K data for the power factor PF, another figure of merit of great interest for thermoelectric applications. Remarkably, the predicted PFs align with the highest values (25/30 $\rm{\mu W/cm\cdot K^2}$) observed in the temperature range of interest for not compatible CMOS material systems (see Refs.~\cite{NatComm2025, PbSe, BiTe}).
PF at this scale  have been recently measured also in Si based devices, as discussed in Ref.~\cite{nanoSi}  where a value of 11 $\rm{\mu W/cm\cdot K^2}$ is reported.
However such high PF requires the nano-patterning of the active region in order to implement an electron energy filter.

\begin{figure}[h!!!]
\centering
\includegraphics[width=0.8\columnwidth]{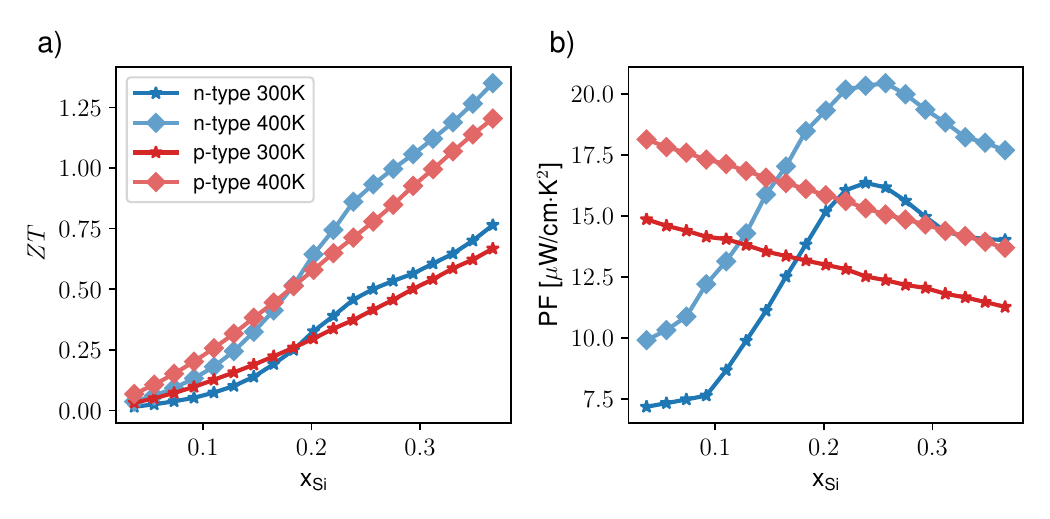}
\caption{ (a) $ZT$ and (b) PF at the Si:Sn ratio of 3.67 as a function of the Si content at 300 and 400 K. The data are for p-type (red) and n-type (blue) alloys, as indicated in the legend.}
\label{fig6}
\end{figure}

Finally, we focus on the special case of interest of SiGeSn alloys with a Si:Sn ratio of 3.67, since in this case the lattice parameter matches the one of Ge~\cite{Ge_match}. The $ZT$ and PF values for this situation are reported in Fig.~\ref{fig6} where a monotonic behavior is observed for the hole based systems  while the n-type material exhibits a more complex behavior due to its multi-valley character. This is particularly apparent in the 0.2-0.3 Si content region where the energy of the $\Gamma$ valley drops down inducing a bump in the $ZT$  and a clear non-monotonic trend in the PF.

\section{Conclusions}
In this article, we studied the thermoelectric properties of the novel CMOS-compatible SiGeSn ternary alloy material system. 
To this aim we have combined lattice thermal conductivity 3-$\omega$ measurements, obtained from state-of-the-art SiGeSn heteroepitaxial layers deposited on Si wafers, with advanced numerical simulations of the carrier transport properties of SiGeSn alloys, performed  taking into account the complex interactions stemming from its multivalley energy band landscape.
The developed full stack of methods and simulation tools has allowed us to numerically explore the experimentally accessible alloy parameter space, delimited by the thermodynamical stability constrains related to the very low solubility of Sn in the Ge and Si lattice.
In n-type systems at the optimal carrier concentration we find a $ZT$ peak value at 300 (400) K of 1.2 (1.6), while the corresponding value for the p-type material is 0.7 (1.3).   
From Fig.~\ref{fig1}(a) it is apparent that these performances are better or in par with those of conventional TE material systems in the same temperature range, typically based on toxic/rare atomic species, which furthermore are hard to impossible to integrate in mainstream electronics.

On this basis our results clearly indicate SiGeSn as a truly promising material system candidate for achieving a new CMOS-compatible TE technology, able to operate efficiently at room temperature. 
We therefore envisage that a transition to a SiGeSn-based TE technology could qualitatively change the application prospects of thermoelectric devices by finally allowing them to be seamlessly integrated into consumer electronics.

\begin{suppinfo}

\begin{itemize}
\item Supplementary Information: material parameters used in the simulations and optimal carrier densities.
\end{itemize}

\end{suppinfo}

\begin{acknowledgement}
M.V., D.M. and S.R. acknowledge support from Italian MUR grant PRIN 2022 PNRR Integrable Thz si-based quantum cascade operation, CUP I53D23006680001 funded by the European Union - NextGenerationEU. P.G. acknowledge the CINECA award under the ISCRA initiative, for the availability of high-performance computing resources and support, and, acknowledges partial funding from the European Union–Next-Generation EU via the Ministry of University and Research of Italy, PRIN PNRR call, grant P20227P7WZ.
G.C., D.B., O.C., J.T-R., A.C-W. acknowledge financial support from the German Research Foundation (DFG) under Projects No. 537127697 “Thermoelektrische Eigenschaften von SiGeSn-Mikrobauelementen”.
\end{acknowledgement}

\bibliography{biblio}

\end{document}